\documentstyle[12pt,titlepage]{article}
\newcommand{\ba}{\begin{array}}
\newcommand{\ea}{\end{array}}
\newcommand{\bd}{\begin{displaymath}}
\newcommand{\ed}{\end{displaymath}}
\newcommand{\be}{\begin{equation}}
\newcommand{\ee}{\end{equation}}
\newcommand{\bea}{\begin{eqnarray}}
\newcommand{\eea}{\end{eqnarray}}

\def\g{\gamma}

\begin{document}
\begin{titlepage}
\vspace*{0.5truein}

\begin{flushright}
\begin{tabular}{l}
MRI-PHY/6/95\\
May, 1995
\end{tabular}
\end{flushright}
\vskip .6cm

\begin{center}
{\Large\bf TAU-UNIVERSALITY VIOLATION WITH \\ LIGHT NEUTRALINOS \\[0.5truein]}
{\large Rathin Adhikari$^1$ and Biswarup Mukhopadhyaya$^2$\\}
Mehta Research Institute \\
10 Kasturba Gandhi Marg\\
Allahabad - 211 002, INDIA \\
\end{center}
\vskip .5cm

\begin{center}
{\bf ABSTRACT}\\
\end{center}

\noindent
In a supersymmetric model with the lightest supersymmetric particle (LSP)
$\chi$ in the range of a few hundred MeV's, the decay
$\tau {\longrightarrow} \mu \chi \chi$ is going to be allowed. We investigate
the departure from tau-universality caused by this decay. It is found that
the universality violation in this way can be greater than both
non-universal electroweak radiative corrections and supersymmetric
one-loop corrections over a considerable region of the parameter space
allowed by experiments so far. Thus it suggests a method of constraining
the parameter space with light LSP's using data from
tau-factories.

\hspace*{\fill}
\vskip .5in

\noindent
$^{1}$E-mail : ~~rathin@mri.ernet.in  \\
$^{2}$E-mail :biswarup@mri.ernet.in

\end{titlepage}

\textheight=8.9in

The search for supersymmetry (SUSY) \cite{susy} has been an active area
for quite some time now. Results from the Large Electron-Positron
(LEP) collider at CERN have put a lower mass bound of $m_{Z}/2$ on most
supersymmetric particles except for the gluino and the lightest supersymmetric
particle (LSP) \cite{pdg}. Also, the Fermilab Tevatron experiments
imply lower bounds
in the range of 150-200 GeV on strongly interacting superparticles like the
squarks and the gluino \cite{tev}. However, it is widely held that
a region in the
parameter space containing light gluinos (2.5-5 GeV) cannot be completely
ruled out yet \cite{lgl}. In such a case, a squark can also be
considerably lighter ($\sim$ 70 GeV or so)  while still evading experiments,
since it will decay
promptly into a quark and a (light) gluino, the latter being instrumental
in degrading the missing transverse energy so that the
corresponding events do not survive the cuts imposed in hadronic collision
experiments. Motivations for a light gluino also come from the observation
that it leads to a better agreement between theory and experiment
in the running of the strong coupling constant $\alpha_s$ \cite{alfas}.
Theoretically,
some attempts have been made in recent times to justify scenarios involving
light gauginos in models of radiative SUSY breaking where dimension-3
terms are absent \cite{far}. Also, efforts have been made to constrain light
superparticles from various phenomenological considerations \cite{sphe}.

Evidently, a light LSP is always present in the scenario described above.
In most SUSY searches, relations among the various parameters
are used to simplify the picture by postulating the SUSY to be embedded
in a Grand Unified Theory (GUT). This has an added incentive in the
demonstration that the three coupling constants can be made to unify
exactly at the energy scale of $10^{16}\;GeV$ if the theory is supersymmetric
\cite{gut}. Under a GUT hypothesis, a light gluino in the range 2.5-5 GeV
normally implies an LSP with mass between 0.4-1.0 GeV. It is also seen that
under such circumstances, all observables are consistent with the LEP
data provided that one is in a region of the parameter space where
$-50\;\; GeV\geq\mu\geq-100\;\;GeV$, and $1.5\leq tan{\beta}\leq 2.0$, $\mu$
and tan$\beta$ being respectively the Higgsino mass parameter and the ratio of
the vacuum expectation values of the two Higgs doublets. Side by side,
some models like those involving radiative SUSY breaking have suggested
LSP's as light as about 100 MeV \cite{far}. It has also been claimed that,
contrary to
earlier conclusions, a light LSP in the range of a few hundred Mev's {\it can}
be reconciled with the dark matter content of the universe \cite{kolb}.
Thus it is desirable to have as many
model-independent criteria as possible to explore a light LSP in the
laboratory. Some such studies have  recently been conducted by us
in the light of B-decay experiments \cite{thap}. Here we would like to
emphasize that the precise measurement of weak universality violation in
$\tau$-decays can also yield useful information in this context.

Weak universality has been found to hold rather accurately in the
$e-\mu$ sector, as is seen from a comparison of the results from pion-decay
with theoretical predictions. Similarly, the universality of charged current
interactions involving the $\tau$ can be subjected to accurate tests in the
decays such as $\tau {\longrightarrow} \mu \overline{\nu_{\mu}} \nu_{\tau}$
and $\tau {\longrightarrow} e \overline{\nu_{e}} \nu_{\tau}$, as also
from $W {\longrightarrow} \tau \overline{\nu_{\tau}}$ and
$\tau {\longrightarrow} \pi \nu_{\tau}$ \cite{marci}.
Precise determination of the mass,
lifetime and the various branching ratios of the $\tau$ in a $\tau$-factory
can further check the standard model predictions in this respect \cite{bill}.

Let us now consider the various ways in which $\tau$-decay may exhibit
departure from universality in the measurements of the leptonic decay
modes. To be specific, let us talk about the decay
$\tau {\longrightarrow} \mu \overline{\nu_{\mu}} \nu_{\tau}$, and
call the corresponding effective Fermi coupling constant $G_{\tau\mu}$.
The total decay width in the above channel, including QED corrections,
is given by \cite{marcsir}

\begin{eqnarray}
\Gamma^{0} = \frac{G^2_{\tau\mu}m^5_{\tau}}{192 \pi^3}\left[1 + \frac{\alpha}
{2 \pi}\left(25/4\;-\; \pi^2\right)\right]\left[1 + \frac{3m^2_{\tau}}
{5m^2_{W}}\right]f(x)
\end{eqnarray}

with

\begin{equation}
f(x) = 1 - 8x +8 x^3 -x^4 -12 \;x^2\; lnx
\end{equation}

The last two factors above correspond to the effects of the $\tau$-momentum
in the W-propagator and the final state muon mass respectively. Here
$G_{\tau\mu}$ is assumed to include the one-loop electroweak radiative
corrections \cite{ewrad} comprising W-boson self-energy, box and triangle
diagrams. Thus $G_{\tau\mu}$ is related to the corresponding quantity
$G_{\mu e}$ by

\begin{equation}
\frac{G_{\tau\mu}}{G_{\mu e}} = 1 + \Delta r_{\tau} - \Delta r_{\mu}
\end{equation}

\noindent where

\begin{equation}
\Delta r_{l} = - \frac{\Pi^T_{WW}(0)}{m^2_W} + box + triangle
\end{equation}

\noindent in the on-shell renormalisation scheme. Thus the deviation of
$\frac{G_{\tau\mu}}{G_{\mu e}}$ from unity depends on the cancellation
between one-loop corrections to $\tau$-and $\mu$-decay, and is of the
order of $\frac{\alpha}{4 \pi} \frac{m^2_\tau}{m^2_W}\;\;\sim\;\;10^{-6}$.

In a SUSY scenario, one-loop graphs involving superparticles further
contribute to $\frac{G_{\tau\mu}}{G_{\mu e}}$. The resulting departures
from universality have been studied in reference \cite{chank} where the
potential contributions from charged-Higgs mediated tree graphs have also been
taken into account. We shall return to comment upon them later.

Our purpose is to point out at this juncture that in the presence of a
light LSP $\chi$, the tree-level decay $\tau {\longrightarrow} \mu \chi \chi$
is also possible. Because of the invisibility of the LSP, this leads to the
same observed final state as $\tau {\longrightarrow} \mu \overline{\nu_{\mu}}
\nu_{\tau}$. As a result, the effective value of $G_{\tau \mu}$ as measured
from $\tau {\longrightarrow} \mu\;+\;nothing$ receives an additional
positive contribution. This contribution is absent in the case of muon decays
if $m_{\chi}\geq m_{\mu}/2$. If we label the width for
$\tau {\longrightarrow} \mu \chi \chi$ as $\Gamma^{SUSY}$, then, neglecting
one-loop effects for the time being, we obtain

\begin{equation}
\frac{G^2_{\tau \mu}}{G^2_{\mu e}} = 1 + \frac{\Gamma^{SUSY}}{\Gamma^0}
\end{equation}

\noindent or

\begin{equation}
\frac{G_{\tau \mu}}{G_{\mu e}} - 1 =\sqrt{1 +
\frac{\Gamma^{SUSY}}{\Gamma^0}}-1
\end{equation}

$\Gamma^{SUSY}$ can receive tree-level contributions because, in a
SUSY model, the lepton and slepton mass matrices are not in general
simultaneously diagonal. This is plausible if one assumes the SUSY to
be embedded in a higher symmetry which is broken at a high energy scale
\cite{evolv}. (One standard way to envision this while at the same time
providing a rather logical method of breaking SUSY is to work with a model
based on N=1 supergravity (SUGRA), the SUGRA being broken at the GUT scale,
leaving as its artifacts soft SUSY breaking terms at the electroweak scale.)
The scalar masses in the resulting theory undergo quantum corrections
as they evolve from the high scale to the electroweak energy.
Thus, if the neutrinos have
non-vanishing masses, the charged slepton mass matrix in the left sector
is given by

\begin{equation}
M^2_{\tilde{l}} = \mu^2  + M_l M_l^{\dagger} + c_0 M_{\nu} M_{\nu}^{\dagger}
\end{equation}

\noindent where the last term arises from to the Yukawa couplings of
left-sleptons with charged
Higgsinos,  $c_0$ being a model-dependent parameter. It is the presence of
this term which causes a mismatch between $M_l$ and $M_{\tilde{l}}$ \
\cite{fcnc}.
Consequently, the lepton-slepton-neutralino interactions in general do violate
flavour. Since the neutrino mass parameters that occur in the Yukawa
couplings correspond to the Dirac mass terms, see-saw type scenarios with
large Majorana masses entail the possibility of such  parameters being of
the order of the tau-mass itself \cite{borz}. Consequently, the
flavour-changing interactions, particularly those involving the third
generation, are also at their strongest in such cases.

The tree-level flavour violating lepton-slepton-LSP coupling
allows the decay $\tau {\longrightarrow} \mu \chi \chi$ through
the diagrams shown in figure 1. With $m_{\chi}$ in the range of a few
hundred MeV's, assuming $\chi$ to be dominantly a photino \cite{arg}, the
flavour-violating interaction is given by

\begin{equation}
{\cal L}_{l_{i}\tilde{l_{j}} \, \chi}  = -{\sqrt{2}} \, e \,c_{ij} \,\,
{\tilde l^{\dagger}_{j}}\,\, {\bar{\chi}} \, \left[\, \,   {{1-\g_5} \over 2}
\right]\,\, {l_i} \, + \, h.c.
\end{equation}

\noindent ${\tilde l}$ being a left slepton. Here $c_{ij}$ is the parameter
characterizing the amount of flavour violation, and is a function of the
parameter $c_0$ and the leptonic mixing matrix. We treat the $c_{ij}$'s as
phenomenological inputs here. The best experimental constraints on them
are obtained from limits on decays like $\mu {\longrightarrow} e \gamma$
and $\tau {\longrightarrow} \mu \gamma$ \cite{mas}. It can be seen by suitably
translating
the limits given in reference \cite{mas} and using the current bounds on
these rare decays \cite{pdg} that while radiative $\mu$-decay leads
to the constraint
$\left(c_{12} {\Delta m^2_{\tilde{l}}}\over{m^2_{\tilde{l}}}\right)_{max}
\approx 10^{-3}$, the restriction on
the $\tau$-sector is much less severe, namely
$\left(c_{23} {\Delta m^2_{\tilde{l}}}\over{m^2_{\tilde{l}}}\right)_{max}
\approx 0.2\;-\;0.3$ (absolute values implied).
For our purpose here the latter is important. Thus, from a
model-independent point of view there is the possibility of relatively
large values of the flavour-changing transition between the third and the
second generations of leptons in a SUSY scenario.

The squared matrix element for
$\tau (p_0) {\longrightarrow} \mu (p_3) \chi (p_1) \chi (p_2)$ is

\begin{eqnarray}
\rule{0.2mm}{3mm}{\cal M}\rule{0.2mm}{3mm}^{2} = \left(\frac{ 64g^{4}Sin^{2}
\theta_{W}c^{2} }{m^4_{\tilde{l}} }\right) \left[ (p_{0}.p_{1}) (p_{2}.p_{3})
+(p_{0}.p_{2}) (p_{1}.p_{3}) - m^2_{\chi}(p_{0}.p_{3})\right]
\end{eqnarray}

\noindent where

\begin{equation}
c =  c_{23} \frac{\Delta m^2_{\tilde{l}}}{m^2_{\tilde{l}}}
\end{equation}

\noindent $m_{\tilde{l}}$ and $\Delta m^2_{\tilde{l}}$ being respectively
the average slepton mass and the mass-squared difference between the left
smu and stau.

The branching ratios for this decay as well as the observed departure
from universality, parametrized by
$\frac{G_{\tau \mu}}{G_{\mu e}} - 1$,
can be directly computed using
equations (6) and (9). Both these quantities are presented as functions of
the LSP mass in figures 2 and 3 respectively. It is obvious from equation (6)
that  to the leading order, $\left(\frac{G_{\tau \mu}}{G_{\mu e}} - 1\right)
\approx (\Gamma^{SUSY}/2\Gamma^{0})\;\;\sim\;\;{c^2}m^{-4}_{\tilde{l}}$.
Thus its dependence on c and the slepton mass can be
studied from Figure 3 itself by suitable scaling.

Figure 2. gives us an idea of the order of magnitude of the branching ratio
for the tau decaying into a pair of LSP's. The curve corresponds to
$m_{\tilde{l}}\;\;=\;\;60\;GeV$ and a 20 per cent slepton mass-squared
splitting. The experimental constraints discussed above allow this region of
the parameter space even upto $c_{23}\;\approx\;1$.

Figure 3. uses two values of the average slepton mass, and
$c\;\;=\;\;0.01$ in magnitude. From the standpoint of experimental limits
this is again
a quite conservative choice of parameters. It is found that the departure
from universality due to $\Gamma^{SUSY}$ can be greater than that from any
other source so long as $m_{\chi}\;\leq\;0.5GeV$,
$m_{\tilde{l}}\;\;\leq\;\;100\;GeV$ and $c\;\leq\;0.1$ approximately.
This immediately suggests the feasibility of limiting a rather large
and hitherto unconstrained area of the parameter space in a scenario with
light LSP's. This should be possible with the accumulation of about
$10^{7-8}$ $\tau$'s in a  $\tau$-factory. The important point to note here
is that the analysis performed here is essentially {\it model-independent} in
nature; even the GUT assumptions are not used. Therefore, any constraints
obtained by this method pertain to non-minimal versions of SUSY as well.

A few comments are in order concerning the one-loop SUSY effects vis-a-vis
the tree-level effects discussed here. Firstly, for the choice of
parameters, if we use the guidelines available from a GUT-inspired scenario,
then a
light LSP (and gluino) should correspond to $tan \beta\;\leq\;2$. Also,
it can be easily verified that the charged Higgs mass has to be about a
hundred GeV so that the LEP limits on the Higgs sector are obeyed. In
such a case, as has been shown in reference \cite{chank}, the Higgs-mediated
one-loop
corrections to $\tau {\longrightarrow} \mu \overline{\nu_{\mu}} \nu_{\tau}$
tend to be small. In a similar way, the Higgs mediated tree-level diagram
gives a very small  (O($10^{-8}$))  contribution  to
$\left(\frac{G_{\tau \mu}}{G_{\mu e}} - 1\right)$. The remaining
part of the one-loop SUSY effect consists in diagrams mediated by
charginos and neutralinos. There, too, the important quantity is
$\Delta r^{SUSY}_{\tau}\;-\;\Delta r^{SUSY}_{\mu}$ {\it i.e.} the
difference between the nonuniversal parts of the two contributions.
The net effect is thus controlled by
$\frac{\Delta m^2_{\tilde{l}}}{m^2_{\tilde{l}}}$ and
$\frac{\Delta m^2_{\tilde{\nu}}}{m^2_{\tilde{\nu}}}$. The enhancements
due to a light LSP in the loop mostly contribute to the universal part
of the correction \cite{uni} and cancel out in
$\left(\frac{G_{\tau \mu}}{G_{\mu e}} - 1\right)$. It is thus estimated
that the loop contributions to leptonic tau-decay with a light
LSP is at best of the order of $10^{-4}$, and that, too, with a rather
large (more than 50 per cent) slepton mass splitting. On the other hand,
our calculations show that the tree level contributions to departure from
universality can be as large as, and perhaps larger than, $10^{-4}$
even for much smaller splitting between
slepton masses. This is evident if one notes that, for example,
$\rule{0.2mm}{3mm}\;c\;\rule{0.2mm}{3mm}\;=\;0.01$
(the value used in Figure 3) is achievable even with
$\rule{0.2mm}{3mm}\frac{\Delta m^2_{\tilde{l}}}{m^2_{\tilde{l}}}
\rule{0.2mm}{3mm}\;=\;0.2$ and $\rule{0.2mm}{3mm}\;c_{23}\;\rule{0.2mm}{3mm}\;
=\;0.05$,
which is well within the region of the parameter space allowed by current
experimental limits. Thus the tree-level flavour-changing decay should give
more useful clues in restricting the parameter space with light LSP's
using departure from $\tau$- universality at the level of reference
\cite{chank}.

A lower limit of about 5 GeV on the LSP mass has been claimed earlier
using the process $e^{+}e^{-} {\longrightarrow} \gamma {\tilde{\gamma}}
{\tilde{\gamma}}$, so long as the selectron is lighter than 55 GeV \cite{asp}.
However, the study of tau decays can improve this limit in a model-independent
manner for either the smuon or the stau having a lower mass.

\newpage
\centerline {\large {\bf Figure Captions}}

\hspace*{\fill}

\hspace*{\fill}

\noindent Figure 1:

\noindent
The tree-level contributions to $\tau \longrightarrow \mu \chi \chi$.
In addition there will be crossed diagrams where the four-momenta of
the LSP's are interchanged.

\vskip .25in

\noindent Figure 2:

\noindent
The branching ratio for  $\tau \longrightarrow \mu \chi \chi$ scaled by
the parameter $c_{23}$, plotted against the LSP mass, for
$\rule{0.2mm}{4mm}\frac{\Delta m^2_{\tilde{l}}}{m^2_{\tilde{l}}}
\rule{0.2mm}{4mm}\;=\;0.2$ and
$m_{\tilde{l}}\;=\;60\;GeV$.

\vskip .25in

\noindent Figure 3:

\noindent
The quantity $\frac{G_{\tau \mu}}{G_{\mu e}} - 1$ plotted against the
LSP mass, for
$\rule{0.2mm}{3mm}\;c\;\rule{0.2mm}{3mm}\;=\;\rule{0.2mm}{5mm}\;c_{23}
\left(\frac{\Delta m^2_{\tilde{l}}}{m^2_{\tilde{l}}}\right)\;
\rule{0.2mm}{5mm}\;=\;0.01$. The solid and dashed lines correspond
to $m_{\tilde{l}}\;=\;45\;GeV$ and $m_{\tilde{l}}\;=\;60\;GeV$ respectively.


\newpage

\begin{thebibliography}{25}

\bibitem{susy} For reviews see, for example,  H. P. Nilles, Phys. Rep.
{\bf 110}, 1 (1984); H. Haber and G. Kane, Phys. Rep. {\bf 117}, 75 (1985).

\bibitem{pdg} Review of Particle Properties, Phys. Rev. {\bf D 50}
(1994).

\bibitem{tev} F. Abe {\it et al.}, Phys. Rev. Lett. {\bf 69 }, 3439
(1992); D. Claes, Talk presented at the 8th meeting of the Division of
Particles and Fields of the American Physical Society, Albuquerque, USA (1994).


\bibitem{lgl}  C. Albajar {\it et al.}, Phys. Lett. {\bf B 198}, 261
(1987); see also [2] and references therein.

\bibitem{alfas}  L. Clavelli, Phys. Rev. {\bf D46}, 2112 (1992); J.
Bl{\"u}mlein and J. Botts, Phys. Lett. {\bf B325}, 190 (1994).

\bibitem{far} G. Farrar and A. Masiero, Rutgers preprint RU-94-38 (1994);
G. Farrar, Rutgers preprint RU-95-17 (1995).

\bibitem{sphe} J. Lopez {\it et al.}, Phys. Lett. {\bf B313}, 241
(1993); M. Diaz, Phys. Rev. Lett. {\bf 73}, 2409 (1994).
C. Carlson and M. Sher, Phys. Rev. Lett. {\bf 72}, 2686 (1994);
G. Bhattacharyya and A. Raychaudhuri, Phys. Rev. {\bf D51}, 2433 (1995).

\bibitem{gut} See, for example, J. Ellis {\it et al.}, Phys. Lett. {\bf B249},
441 (1990), {\it ibid}, {\bf B260}, 131 (1991); U. Amaldi {\it et al.},
Phys. Lett. {\bf B260}, 447 (1991); P. Langacker and M. Luo, Phys. Rev. {\bf
D44}, 817 (1991); G. Ross and R. Roberts, Nucl. Phys. {\bf B377}, 571 (1992).

\bibitem{kolb} G. Farrar and E. Kolb, Fermilab preprint FERMILAB-PUB-95/068-A
(1995).

\bibitem{thap} R. Adhikari and B. Mukhopadhyaya, MRI preprint MRI-PHY/19/94
(1994), MRI-PHY/21/94 (1994).

\bibitem{marci} W. Marciano and A. Sirlin, Phys. Rev. Lett. {\bf 61}, 1815
(1988); W. Marciano, Phys. Rev. {\bf D45}, R721 (1992); D. Bryman, Phys. Rev.
{\bf D46}, 1064 (1992).

\bibitem{bill} W. Marciano, Talk Presented at the Third Workshop on Tau
Lepton Physics, Montreux, Switzerland (September, 1994).

\bibitem{marcsir} W. Marciano and A. Sirlin, Phys. Rev. Lett. {\bf 71},
3629 (1988).

\bibitem{ewrad} W. Marciano and A. Sirlin, Phys. Rev. {\bf D22}, 2695 (1980);
W. Hollik, Fort. Phys. {\bf 38}, 165 (1990).

\bibitem{chank} P. Chankowski {\it et al.}, Phys. Lett. {\bf B333}, 403 (1994).

\bibitem{evolv} R. Barbieri {\it et al.}, Phys. Lett. {\bf B119}, 343 (1982);
A. Chammesdine {\it et al.}, Phys. Rev. Lett. {\bf 49}, 970 (1982); L. Hall
{\it et al.}, Nucl. Phys. {\bf B267}, 451 (1986).

\bibitem{fcnc}   J. Donoghue {\it et al.}, Phys. Lett. {\bf B128}, 55 (1983);
M. Duncan, Nucl. Phys. {\bf B221}, 285 (1993); A. Bouquet {\it et al.},
Phys. Lett. {\bf B148}, 69 (1984).

\bibitem{borz} F. Borzumati and A. Masiero, Phys. Rev. Lett. {\bf 57}, 961
(1986); B. Mukhopadhyaya and A. Raychaudhuri, Phys. Rev. {\bf D42}, 3215
(1990).

\bibitem{arg} The region of the parameter space to which one has to confine
oneself to simultaneously accommodate a light gluino and and the LEP data
leads to a neutralino mass marix where the LSP is almost entirely the photino.
We have used this as a guiding principle although otherwise our study does
not require the GUT assumptions.

\bibitem{mas} F. Gabbiani and A. Masiero, Nucl. Phys. {\bf B322}, 235 (1989).

\bibitem{uni} P. Chankowski {\it et al.}, Nucl. Phys. {\bf B417}, 101 (1994).

\bibitem{asp} C. Hearty {\it et al.}, Phys. Rev. {\bf D39}, 3207 (1989).

\end{thebibliography}
\end{document}